\documentclass[preprint,aps,showpacs,prb]{revtex4}

\usepackage{amsfonts}

 \usepackage{graphicx}
 \usepackage{amsmath}
 \usepackage{natbib}
 \usepackage{epsfig}

\begin{document}
\title{ Spin accumulation in ferromagnets}
\author{W. N. G. Hitchon$^{1}$}
\author{R. W. Chantrell$^{2}$}
\author{A. Rebei$^{2}$ \footnote{Corresponding
author:arebei@mailaps.org}}
 
\affiliation{
   $^{1}$Department of Electrical and Computer
Engineering, University
of Wisconsin-Madison, Madison, Wisconsin 53706, USA\\
$^{2}$Seagate Research Center, Pittsburgh, Pennsylvania 15222,
USA
}
  
\date{06-30-04 }

\begin{abstract}
Using a density matrix formulation for the effective action, we obtain a set
of macroscopic 
equations that describe the spin accumulation in a non-homogeneous
ferromagnet. We give a new expression for the spin current which extends
previous work by 
taking into account the symmetry of 
the ferromagnetic state through a careful treatment of the 
exchange term between the conduction electrons and the 
magnetization, i.e., d-electrons. 
 We consider a simple application which has been discussed 
previously and 
show that in this case spin accumulation is an interface 
effect confirming earlier results arrived at by different 
methods. 
\end{abstract}

\pacs{76.30.Pk, 72.25.-b, 75.10.-b}
\maketitle

Spin-momentum transfer (SMT), was predicted by Berger \cite{Berger}
and Slonczewski \cite{Slon} and observed experimentally \cite{Cornell,Zurich}
and studied theoretically by various groups. \cite
{Heidi,BAZALIY,Waintal,Zhang,Stiles,Bauer,Simanek,Mills,Ho}. The recent work 
by Zhang, Levy and
Fert (ZLF) \cite{Zhang} argued that besides the spin torque, the sd-exchange
term between the conduction electrons and the magnetization gives rise to an
extra effective field which also can induce switching by precession. They
showed that its effect can be important to distances larger
than $1.0 nm$
from the interface. \ Calculations by Stiles and 
Zangwill \cite{Stiles} and by
Slonczewski \cite{Slon} show indirectly that the effect of this field is
almost absent. \ These calculations give the impression that a macroscopic 
 treatment, such as the ZLF calculation, is either
not applicable or simply gives a different result for 
some unknown reasons. \ However a macroscopic approach is very appealing
since it
 can 
easily be integrated within micromagnetics and keeps the problem of spin
accumulation within the reach of classical methods. \ In fact as we show
elsewhere \cite{chantrell}, the treatment of spin accumulation through a
spin density vector provides the most straightforward extension of the
Valet-Fert theory in CPP structures to non-collinear 
configurations of the local magnetization.\cite{valet} \ We integrate
the Boltzmann equation to obtain macroscopic equations of motion 
for the spin accumulation.\ In this communication, we show how to 
extend the ZLF treatment 
 and give generalized equations for the spin diffusion in the
presence of non-uniform magnetization which may be useful for problems
involving domain walls. 

As an application of our results, we
show that  in the uniform magnetization case, the effective 
field 
predicted by ZLF
is 
indeed
vanishingly small for distances larger than $1.0 nm$ from the F/N interface.
This is in agreement with the other calculations mentioned above.\cite{Slon,
Stiles} \  Hence, at
least in the uniform case, the macroscopic treatment also predicts that SMT
 is a surface effect. \ Our equations differ from those
derived previously by ZLF by including the effect of the magnetization on
the electron propagators at the microscopic level. \  We start from a
microscopic description of the conduction electrons and the ferromagnetic
medium and take the semi-classical limit to derive equations for macroscopic
quantities in the 
diffusive limit, generalizing 
previous results \cite{Heidi,Zhang}. \ We believe that this generalization 
is necessary for metallic elements.  \ Details of
the derivation which is based on a path integral 
approach are treated elsewhere. \cite{Hitchon}

\bigskip

We  first introduce the notation: With a spin vector $\mathbf{S}_{i}$ at each
lattice point $i$, the macroscopic spin vector for the medium is 
\begin{equation}
\mathbf{S}\left( \mathbf{r}\right) =\sum_{i=1}^{N}\mathbf{S}_{i}\delta\left( 
\mathbf{r}-\mathbf{r}_{i}\right) \;.
\end{equation}
 The interaction between the electrons and the localized spins is
taken to be of the s-d type, of the form 
\begin{equation}
H_{sd}=-\frac{{J}}{2}\int d\mathbf{x}\left( \mathbf{\Psi}^{\dagger}\left( 
\mathbf{x}\right) \mathbf{\overrightarrow{\mathbf{\sigma}}}\mathbf{\Psi }%
\left( \mathbf{x}\right) \right) \cdot\mathbf{S}\left( \mathbf{x}\right)
\end{equation}
where $J$ is a coupling constant of the order of $0.1\;eV$ and $%
\overrightarrow{\mathbf{\sigma}}$ is a vector whose components are the Pauli
matrices, 
\begin{equation}
\left[ \sigma_{i},\sigma_{j}\right] =2i\epsilon^{ijk}\sigma_{k}.
\end{equation}
$\epsilon^{ijk}$ is the antisymmetric unit tensor. \noindent The Hamiltonian
of the theory has the form 
\begin{align}
\mathcal{H} & =\sum_{\alpha=1,2}\int d\mathbf{r}\;\Psi_{\alpha}^{\ast}\left( 
\mathbf{r}\right) \left\{ \frac{1}{2m^{\ast}}\mathbf{p}^{2}\left( \mathbf{r}%
\right) +U\left( \mathbf{r}\right) -\mathbf{\sigma}\cdot \mathbf{B}\right\}
\Psi_{\alpha}\left( \mathbf{r}\right) \\
& -\frac{{J}}{2}\int d\mathbf{r}\sum_{\alpha,\beta=1}^{2}\sum_{i=1}^{3}%
\Psi_{\alpha}^{\ast}\left( \mathbf{r}\right) \sigma_{\alpha\beta}^{\left(
i\right) }\Psi_{\beta}\left( \mathbf{r}\right) {S}_{i}\left( \mathbf{r}%
\right) +\mathcal{H}_{M}  \notag
\end{align}
where $\mathbf{B}$ is an external magnetic field, $U\left( r\right) $ is \ a
spin-independent potential that includes the electric field and $\mathcal{H}%
_{M}$ is the Hamiltonian of the magnetic system alone without the conduction
electrons. \ In the above, 
we are using units such that $\hbar = g \mu_{B} = 1$.
\ We use methods of non-equilibrium field theory to extract the
semi-classical equations of motion for the spin accumulation 
$\mathbf{m}$. \cite{Schwinger,Haug}
Our treatment is similar in spirit to that of Brataas, Nazarov and Bauer \cite{Bauer} except 
that here 
we use the sd-exchange model to simulate the spin momentum exchange between the 
electrons and the magnetization. Recently Mills \cite{Mills} and before that Berger \cite{Berger}
gave a detailed treatment of this exchange in the ballistic regime. Here we focus on the diffusive
regime which is applicable in large devices such as CPP recording heads 
(see Ref. \onlinecite{Bauer} for 
a further discussion of this point). \ To get a macroscopic description of spin 
accumulation, we need to derive 
equations of motion for the   
 two-point propagators of the conduction electrons' field  
\begin{align}
\mathcal{G}_{22}^{ss^{\prime}}\left( \mathbf{x},\mathbf{y}\right) &
=\langle\;\mathcal{T}^{-1}\left( \Psi_{s}\left( \mathbf{x}\right)
\Psi_{s^{\prime}}^{+}\left( \mathbf{y}\right) \;\right) \rangle, \\
\mathcal{G}_{21}^{ss^{\prime}}\left( \mathbf{x},\mathbf{y}\right) &
=\langle\Psi_{s}\left( \mathbf{x}\right) \Psi_{s^{\prime}}^{+}\left( \mathbf{%
y}\right) \rangle,  \notag \\
\mathcal{G}_{11}^{ss^{\prime}}\left( \mathbf{x},\mathbf{y}\right) &
=\langle\;\mathcal{T}\left( \Psi_{s}\left( \mathbf{x}\right) \Psi
_{s^{\prime}}^{+}\left( \mathbf{y}\right) \right) \;\rangle,  \notag \\
\mathcal{G}_{12}^{ss^{\prime}}\left( \mathbf{x},\mathbf{y}\right) &
=-\langle\Psi_{s^{\prime}}^{+}\left( \mathbf{y}\right) \Psi_{s}\left( 
\mathbf{x}\right) \rangle.  \notag
\end{align}
$\mathcal{T}$ is the usual time ordering operator and $\mathbf{x}$ is 
a 4-vector with time and spatial components.

Using the standard tools of field theory we derive the approximate effective
action for the conduction electrons and the localized magnetic moments (see Ref. 
\onlinecite{Hitchon} and references therein for more details). \  From
the effective action we obtain equations of motion for the spin current and
the magnetic moment. We use the true electron propagator taking into account
the local magnetization in a self-consistent way.\ This self-consistent 
treatment of the effect of the magnetization on the electrons 
 is the main goal of this 
work. \ We define

\begin{align}
\mathfrak{M}_{\alpha\beta}^{i}\left( \mathbf{x},\mathbf{y}\right) \; & = \;%
\frac{1}{2}\sigma_{ss^{\prime}}^{i} \mathcal{G}_{\alpha\beta}^{s^{\prime}s}%
\left( \mathbf{x},\mathbf{y}\right) \;,
\end{align}
to be the conduction electron spin propagator. The classical polarization of
the current $\mathbf{m}$ is found by averaging over the fast degrees of
freedom, in the center of mass coordinates and using the quasi-particle
approximation. \cite{Haug}

The spin current $\mathbf{\mathfrak{J}}$, a fourth rank tensor,
 is defined in the usual way 
\begin{equation}
\mathfrak{J}^{kl}\left( t,\mathbf{x} \right) =\frac{V}{m^{\ast}}
\int\frac{d\mathbf{p}}{%
(2\pi)^{3}}\;\mathfrak{M}^{k}\left( t,\mathbf{x},\mathbf{p}\right) p^{l},
\end{equation}
where $t$ and $\mathbf{x}$ are now coordinates in the center of mass. \ 
The spin accumulation vector is therefore given by 
\begin{equation}
m^{k}\left( t,\mathbf{x}\right) =\frac{V}{m^{\ast}}\int\frac{d\mathbf{p}}{(2\pi)^{3}}%
\;\left[ \mathfrak{M}^{k}\left( t,\mathbf{x},\mathbf{p}\right) -\mathfrak{M}_{eq}^{k}\left(
\mathbf{x},\mathbf{p}\right) \right] .
\end{equation}

To simplify the treatment, we assume that the splitting of the conduction
electron bands is small with respect to the Fermi 
energy, i.e., the polarization 
of the conduction electrons is small. \ We have also used a 
quasiparticle approximation for the collision term.
\ For slow variations in time, we have a modified Fick's law and Ohm's law for 
the spin accumulation 
\begin{equation}
\mathfrak{J}_{j}^{k}\left( t,\mathbf{x}\right) =-\gamma m_{eq}^{k}\left(
\mathbf{x}\right) E_{j}-\mathfrak{D}^{kp}\partial_{X_{j}}m^{p}\left( t,\mathbf{x}%
\right) ,   \label{flux}
\end{equation}
where $\gamma$ is the effective mobility \cite{Heidi}. \ This is the major 
result of this communication which improves on previous expressions 
used for spin accumulation in ferromagnets and  is 
valid for a general configuration of the local 
magnetization. \ The effective diffusion
tensor is  
\begin{equation}
\mathfrak{D}^{kp}\left( \mathbf{x}\right) =D_{\parallel}\left( A^{-1}\right) ^{kp}(%
\mathbf{x}),
\end{equation}
where $D_{\parallel}$ is the diffusion coefficient and 

\begin{equation}
A\left( \mathbf{\mathbf{x}}\right) =\left[ 
\begin{array}{ccc}
1 & -\tau H_{z} & \tau H_{y} \\ 
\tau H_{z} & 1 & -\tau H_{x} \\ 
-\tau H_{y} & \tau H_{x} & 1
\end{array}
\right] 
\end{equation}
with the effective local field given by

\begin{equation}
\mathbf{H}=\mathbf{B}+J \, \mathbf{M}\left( \mathbf{x}\right) .
\end{equation}
$\tau$ is the (momentum) relaxation time. \ The diffusion coefficient in Eq.
 \ref{flux} is a tensor with symmetry about the 
  local axis of the magnetization 
  whereas  in ZLF the diffusion coefficient is a scalar.  \ This will have important 
consequences as shown below. \ In steady state, the 
average magnetic moment $\mathbf{m}$ obeys 
\begin{equation}
\sum_{p,l}\partial_{x_{l}}\left[ \mathfrak{D}^{kp}\left( x\right)
\partial_{x_{l}}m^{p}\left( \mathbf{x}\right) \right] =\frac{1}{\tau_{sf}}%
m^{k}\left( \mathbf{x} \right) + J \left[ \mathbf{M}\left( \mathbf{x} \right)
\times \mathbf{m}\left( \mathbf{x} \right)
   \right] ^{k}.   \label{diffusion}
\end{equation}

\ Next, we apply this result to 
a case similar to that in ZLF. \  First, we write explicitly Eq. \ref{diffusion}
  for a  magnetization which is a function only 
of distance x in the direction of the
current, $\mathbf{M}=M\left( x\right) \mathbf{z}$,

\begin{align}
D_{p}\frac{d^{2}m_{x}\left( x\right) }{dx^{2}}+D_{xy}\frac{d^{2}m_{y}\left(
x\right) }{dx^{2}} &  \notag \\
-2\frac{D_{xy}^{2}}{Da\left( x\right) }\frac{da\left( x\right) }{dx}\frac{%
dm_{x}\left( x\right) }{dx}+\left( D_{p}-2\frac{D_{xy}^{2}}{D}\right) \frac{%
da\left( x\right) }{dx}\frac{dm_{y}\left( x\right) }{dx} & =\frac{
m_{x}\left( x\right) }{\tau_{sf}}-\frac{a\left( x\right) m_{y}\left( x\right) }{%
\tau_{sf}}
\end{align}
\begin{align}
-D_{xy}\frac{d^{2}m_{x}\left( x\right) }{dx^{2}}+D_{p}\frac{d^{2}m_{y}\left(
x\right) }{dx^{2}} &  \notag \\
-\left( D_{p}-2\frac{D_{xy}^{2}}{D}\right) \frac{da\left( x\right) }{dx}%
\frac{dm_{x}\left( x\right) }{dx}-2\frac{D_{xy}^{2}}{Da\left( x\right) }%
\frac{da\left( x\right) }{dx}\frac{dm_{y}\left( x\right) }{dx} & =\frac{%
m_{y}\left( x\right) }{\tau_{sf}}+\frac{a\left( x\right) m_{x}\left( x\right) }{%
\tau_{sf}}
\end{align}
where
\begin{equation}
a\left( x\right) =\tau J M\left( x\right) ,
\end{equation}
and the remaining coefficients are given below. \
If the magnetization is a function of x, y and z, the equations are even
more involved. In this case,  the transverse components of the spin
accumulation will not decouple from the components along the magnetization 
\cite{Hitchon}. \ As a consequence, the spin accumulation can be enhanced by
choosing appropriate gradients of the magnetization along the interfaces. \ If
 we assume that the local magnetization is uniform and along the z-axis, the 
equations simplify considerably and 
we find (instead of Eqs. 11 and 13 of Ref.~\onlinecite{Zhang}).

\begin{equation}
D_{p}\frac{d^{2}}{dx^{2}}m_{x}\left( x\right) + D_{xy}\frac{d^{2}}{dx^{2}}%
m_{y}\left( x\right) =\frac{m_{x}}{\tau_{sf}}- J M m_{y}\left( x\right)
\end{equation}
\begin{equation}
-D_{xy}\frac{d^{2}m_{x}\left( x\right) }{dx^{2}}+D_{p}\frac{d^{2}}{dx^{2}}%
m_{y}\left( x\right) =\frac{m_{y}}{\tau_{sf}}+J M m_{x}\left( x\right)
\end{equation}
\begin{equation}
D_{\parallel}\frac{d^{2}m_{z}\left( x\right) }{dx^{2}}=
\frac{m_{z}}{\tau_{sf}}
\end{equation}
with the diffusion coefficients defined by 
\begin{equation}
D_{p}=D_{xx}=D_{yy}=\frac{D_{\parallel}}{1+\left( \tau J M(x)\right) ^{2}}
\end{equation}
and the off-diagonal terms, which do not appear in the ZLF theory, are
\begin{equation}
D_{xy}=-D_{yx}=D_{\parallel}\frac{\tau J M(x)}{1+\left( \tau J M(x)\right) ^{2}}.
\end{equation}

In transition metals, the coefficient  $\left| D_{xy}\right|$ $\sim $ 100 $D_{p}$ (in the bulk) and $D_{\parallel }$ is
about three orders of magnitude larger than $D_{xy}$ for a transition metal such as nickel.\cite{Hirst}\  These equations are
strictly valid for the bulk. Any account of the N/F interface is taken care
of by suitable boundary conditions \cite{valet,chantrell}. The
equations are given in their simplest form to compare to Ref.~\onlinecite{Zhang},
which similarly does not take account of the interface. The off-diagonal
terms are absent in all previous works on spin-momentum transfer. However,
the above equations are similar to equations derived by Hirst ~\cite
{Hirst} and Kaplan ~\cite{Kaplan} for a circularly polarized itinerant
electron gas. In Refs.~\onlinecite{Hirst} and 
\onlinecite{Kaplan} the off-diagonal terms are attributed
to exchange stiffness. In our case, the non-diagonal terms $D_{xy}$ are also
due to an effective exchange between the conduction electrons mediated by
the magnetization $\mathbf{M}$ along the $z-$axis. \ It is these 
exchange coefficients that sets the scale of spin transfer in the 
problem when $\tau J M >> 1$. 
The equations now reflect
the symmetry of the ferromagnetic state as required. The different terms
presented here will have a significant effect on the Valet-Fert theory when
extended to treat non-collinear magnetizaton in the different layers \cite
{chantrell} and on the result reported in ZLF about the effectiveness of the 
ZLF mechanism, i.e., the transverse field. \ The symmetry is easily seen by introducing the
complex spin accumulation $\mathfrak{m}$,
\begin{equation}
\mathfrak{m}\left( x\right) =m_{x}\left( x\right) - i m_{y}\left( x\right) 
\end{equation}
a complex diffusion coefficient $D_{c}$,
\begin{equation}
D_{c}=D_{p } + i D_{xy},
\end{equation}
and a complex relaxation time $\tau _{c}$,
\begin{equation}
\frac{1}{\tau _{c}}=\frac{1}{\tau_{sf} } + i J M(x).
\end{equation}
The steady-state transverse spin accumulation obeys,
\begin{equation}
\frac{d^{2}\mathfrak{m}}{dx^{2}}=\frac{\mathfrak{m}}{\lambda _{c}^{2}},
\end{equation}
where
\begin{equation}
\lambda _{c}^{2}=\tau _{c}D_{c}.
\end{equation}
The local z-component of the spin accumulation obeys
\begin{equation}
\frac{d^{2}m_{z}}{dx^{2}}=\frac{m_{z}}{\lambda _{sdl}^{2}},
\end{equation}
where $\lambda _{sdl}$ is the (longitudinal) spin diffusion length which can
be 5-100 nm. The general solutions for the complex accumulation are of the
form:
\begin{equation}
\mathfrak{m}\left( x\right) =A\exp \left[ -x/\lambda _{c}\right] +B\exp
[x/\lambda _{c}].
\end{equation}
The spin accumulation therefore shows 
in general an exponential decrease (or increase) 
with some oscillations. 
In this notation, the spin current is, omitting the electric field, given by
two components, the perpendicular one which is complex and the longitudinal
one, 
\begin{align}
j_{\perp }& =-D_{c}\frac{d\mathfrak{m}}{dx}, \\
j_{z}& =-D_{\parallel }\frac{dm_{z}}{dx}.
\end{align}

\  In Ref.~\onlinecite{Zhang}, the Landau-Lifshitz equation is

\begin{equation}
\frac{d\mathbf{M}}{dt}=-\gamma _{0}\mathbf{M}\times \left( \mathbf{H}_{eff}+b%
\mathbf{M}_{REF}\right) -\gamma _{0}a\mathbf{M}\times \left( \mathbf{M}%
_{REF}\times \mathbf{M}\right) +\alpha \mathbf{M\times }\frac{d\mathbf{M}}{dt%
}
\end{equation}
$a$ and $b$ measure the relative strength of the Slonczewski term and the
transverse term, respectively. In ZLF~\cite{Zhang}, the ratio $b/a$ was
found to be almost 2 for their 
 $\lambda _{J}=2.0$ nm, the transverse spin accumulation 
length. \ To avoid some problems with
the boundary conditions in the solution of ZLF, we solve the same problem
for a semi-infinite ferromagnet using the ZLF equations and our 
equations for the {\it same} parameters. \ We assume, as they did,  the spin 
current
is initially polarized  along the 
vector $\mathbf{u}=\left( 0,-\sin \theta ,\cos
\theta \right) $ by another layer with a magnetization parallel to $\mathbf{%
u}$. Figure 1 shows the original solution of ZLF for the
x and y components. \ In 
 Fig.2 we show our solution for the same boundary conditions, i.e., 
we assume that the spin current is continuous and the transverse 
components of the spin accumulation decay to zero in the 
ferromagnet. \ The spin 
diffusion length is taken to be 
the same in both cases. \ Both figures are given for $\lambda = 10.0$ nm,  
$J = 0.10$ eV and $\tau_{sf} = 10^(-11)$ s which are typical parameters 
of a soft permalloy material.
 \ With these parameters, the spin 
accumulation in the ZLF case 
seems to be appreciable for distances up to almost $10 nm$, while in our 
case it vanishes within $1.0 nm$ from the interface.\ Our result 
is in agreement with other 
calculations \cite{Stiles} and
 clearly shows that 
spin accumulation is a surface
effect. \ This can be traced to the fact that 
the polarization of the conduction electrons in a 
ferromagnet diffuse much more slowly in the 
direction perpendicular to the magnetization as opposed to 
that along the magnetization due to the s-d exchange in our 
case. \ It should also 
be observed that 
in the limit where spin-flip scattering is important,
 the  oscillations in the spin accumulation are ``overdamped''. \ This is in 
contrast to what 
 the quantum calculations of Berger \cite{Berger}, 
Mills \cite{Mills} and in Ref. \onlinecite{rebei00} give which 
does not take into account spin relaxation effects. \ These 
latter oscillations are probably an artifact of the free electron 
band model as shown recently in Ref.\onlinecite{Bauer2}.

In summary, we have written the equations for 
 the spin accumulation in a non-homogeneous ferromagnet 
in a form which reflects the 
symmetry of the state and obtained a non-trivial self-consistent 
expression for Fick's law. \ We find that the predictions of the 
  ZLF  theory  are correct only if we neglect exchange 
effects. \ This is clearly not the case in a transition metal. \ We have 
shown that a correct inclusion of exchange effects greatly constrain 
the spin accumulation to be important only near to the surface. \ This 
conclusion
 may not be valid
  when the magnetization
is a function of all three coordinates.

\bigskip

\begin{figure}[h]
\begin{center}
\mbox{\epsfig{file=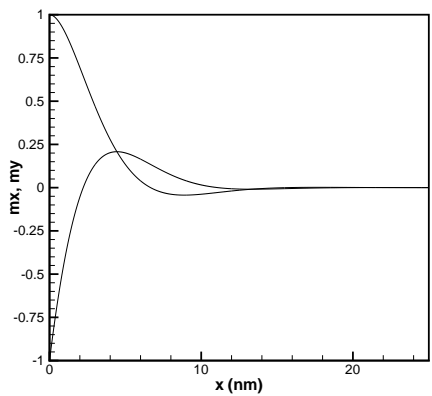,height=6 cm}}
\end{center}
\caption{ZLF solution for a half-plane geometry. The x and y components are normalized to one.}
\end{figure}

\begin{figure}[h]
\begin{center}
\mbox{\epsfig{file=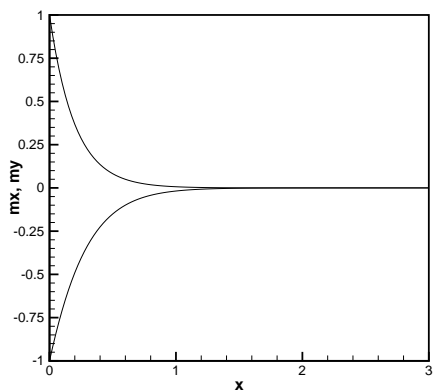,height=6 cm}}
\end{center}
\caption{Our solution for a half-plane geometry for the same 
parameters as in Fig.1 (see text). The x and y components are normalized to one. Note the difference in scale along the x-axis (nm) with respect to fig.1.}
\end{figure}

%
%

\end{document}